\documentstyle[11pt,amssymbols,epsf]{article}
\newcommand{\be}{\begin{equation}}
\newcommand{\ee}{\end{equation}}
\newcommand{\ba}{\begin{array}}
\newcommand{\ea}{\end{array}}
\newcommand{\bea}{\begin{eqnarray}}
\newcommand{\eea}{\end{eqnarray}}

\renewcommand{\l}{\newline\null}

\newcommand{\rar}{\rightarrow}
\newcommand{\p}{\partial}
\newcommand{\ol}{\overline}
\newcommand{\ti}{\tilde}
\newcommand{\la}{\langle}
\newcommand{\ra}{\rangle}
\def\figskip{\vskip .5cm plus 3mm minus 2mm}

\begin{document}
\begin{titlepage}
June 1993\hfill PAR-LPTHE 93/35
\begin{flushright} hep-ph/9306285 \end{flushright}
\vskip 4cm
\begin{center}
{\bf  THE COUPLINGS OF THE  PION TO TWO GAUGE FIELDS AND TO LEPTONS \\
IN A DYNAMICALLY BROKEN GAUGE THEORY}
\end{center}
\vskip 1cm
\centerline{B. Machet
     \footnote[1]{Member of Centre National de la Recherche Scientifique}
     }
\vskip 5mm
\centerline{{\em Laboratoire de Physique Th\'eorique et Hautes Energies}
     \footnote[2]{LPTHE tour 16/1er \'etage,
          Universit\'e P. et M. Curie, BP 126, 4 place Jussieu,
          F 75252 PARIS CEDEX 05 (France).}
     \footnote[3]{E-mail:  machet@lpthe.jussieu.fr},
                                                          {\em Paris}}
\centerline{\em Unit\'e associ\'ee au CNRS URA 280.}
\vskip 2cm
{\bf Abstract:}  we show  how a spontaneously broken gauge theory of fermions
endowed with a composite scalar multiplet becomes naturally anomaly-free, and
yet describes the correct couplings of the pion to two gauge fields and to
leptons: the first coupling is the same as computed from the chiral anomaly,
and the second identical with that obtained from the `Partially Conserved
Axial Current' hypothesis. For the sake of simplicity, we only study here
the abelian case.
\smallskip

{\bf PACS:} \quad 11.15.Ex, 11.40.Fy, 11.40.Ha, 12.15.Cc, 12.50.Lr, 13.20.Cz,
13.40.Hq, 14.40.Aq, 14.80.Gt
\vfill
\null\hfil\epsffile{LOGO.eps}
\end{titlepage}

\section{Introduction}

Spontaneously broken gauge theories, embodied by the Standard Model of
electroweak interactions \cite{GlashowSalamWeinberg}, face two major issues:\l
\quad$\ast$\  anomalies \cite{AdlerBellJackiwBardeen}:
gauge anomalies are an obstacle to renormalizability and are required to
cancel between quarks and leptons \cite{BouchiatIliopoulosMeyer};
chiral anomalies yield however  the correct decay of the neutral pion into
two photons \cite{Adler}.\l
\quad$\ast$\  the lack of prediction concerning their scalar sector.

Considering the latter as composite went along, up to now, with the
introduction of another scale of interaction \cite{SusskindWeinberg}, but
`technicolour' theories run into serious problems \cite{FarhiJackiw}. This
has led to the blockage that we are facing nowadays, worsened by the growing
feeling that supersymmetry may have nothing to do with nature \cite{Wess}.

We present here an abelian dynamically broken gauge theory which provides the
correct coupling of the (neutral) pion to leptons and, though being anomaly
free, also yields its usual coupling to two gauge fields. The non-abelian
case is included in the more general work \cite{BellonMachet2}.
We do not worry either about renormalizability and only mention the arguments
in favour of it, which are developed in
\cite{BellonMachet2,Machet2,BellonMachet3}.

This work is completed by \cite{BellonMachet1} where we show how the
Standard Model of leptons can be reconciled with an anomaly-free purely
vectorial theory, thus cutting the link between the hadronic and leptonic
sectors \cite{BouchiatIliopoulosMeyer}.

\section{The hadronic sector}

We consider a $U(1)_L$ spontaneously broken gauge theory; the generator of
the gauge group $\cal G$ is
\be
{\Bbb T}_L = {1-\gamma_5\over 2}\; {\Bbb T};
\ee
it acts on the gauge field $\sigma_\mu$ and on the $N$ fermions.
$\Psi$ is the $N$-vector: $N$ is the number of `flavours'.
we embed ${\cal G}$ into the chiral group $U(N)_L \times U(N)_R$ and consider
$\Bbb T$ as a $N\times N$ matrix.

When ${\Bbb T}^2 = 1$
\be
\Phi =(H,\varphi) = {v\over\mu^3}(\ol\Psi\Psi, -i\ol\Psi\gamma_5{\Bbb T}\Psi)
\ee
is a 2-dimensional representation of the gauge group
(this condition is generalized in the Standard Model to the existence of an
associative algebra \cite{BellonMachet2}). Both $H$ and $\varphi$  are real.
$H$ will be called the Higgs boson by analogy. We take for example in the
following ${\Bbb T}=1$, and, from the action of $\cal G$ on the
fermions, we then deduce
\be\left\{\ba{lcl}
                {\Bbb T}_L.\varphi &=& iH,\cr
                {\Bbb T}_L.H    &=& -i\varphi. \ea \right .
\ee
The gauge symmetry is spontaneously broken by $\la H\ra = v$, equivalent to
$\la\ol\Psi\ol\Psi\ra = \mu^3$. The gauge and chiral symmetry breaking are
thus two aspects of the same phenomenon. We write as usual
\be
H = v+ h.
\ee
\pagebreak

The Lagrangian is chosen as
\bea
{\cal L} = -{1\over 4}F_{\mu\nu}F^{\mu\nu}
&+& i \ol\Psi\gamma^\mu (\p_\mu -ig\;\sigma_\mu \, {\Bbb T}_L )\Psi
+{1\over 2}(D_\mu H D^\mu H + D_\mu\varphi D^\mu \varphi) -V(H^2 +
\varphi^2)\cr
&-& {\p_\mu\xi\over v}\, \ol\Psi\gamma^\mu\,{\Bbb T}_L\Psi,
\label{eq:L}\eea
where the real field
\be
\xi = -\varphi\; (1- {h\over v}) + \cdots
\ee
is defined, together with
\be
\tilde H = v+ \eta
\ee
by
\be
\tilde H  = e^{-i{\xi\over v}{\Bbb T}_L}.\,(H + i\varphi)
\ee
When
\be
\Psi \rar e^{-i\theta\,{\Bbb T}_L}\, \Psi,
\ee
it transforms, like a $U(1)$ Wess-Zumino \cite{WessZumino} field, by
\be
\xi \rar \xi - \theta v
\ee
while $\tilde H$ stays invariant. One has $\ti H^2 = H^2 + \varphi^2$.
$\cal L$  in equation (\ref{eq:L}) differs from the `standard' Lagrangian by
the additional coupling
\be
- {\p_\mu\, \xi\over v}\; \ol\Psi \gamma^\mu{\Bbb T}_L \Psi;
\label{eq:xicpl}\ee
its presence is  however the natural consequence of taking $\cal L$ as a
function of the `generic' gauge field $\sigma_\mu - (1/ g)\p_\mu \xi/v$ instead
of $\sigma_\mu$ alone; indeed, $\cal L$ can also be written (see for example
\cite{AbersLee})
\bea
{\cal L} &=& -{1\over 4}F_{\mu\nu}F^{\mu\nu}
+ i \ol\Psi\gamma^\mu \Big(\p_\mu -ig\;(\sigma_\mu -
{1\over g}{\p_\mu \xi\over v})\; {\Bbb T}_L \Big)\Psi\cr
& &+{1\over 2}(\p_\mu \ti H)^2
+ {1\over 2} g^2\Big(\sigma _\mu -{1\over g}{\p_\mu \xi\over v}\Big)^2
\tilde H^2 - V(\tilde H^2).
\eea
This has been advocated in \cite{BabelonSchaposnikViallet}
to lead to the recovery of gauge invariance for anomalous gauge theories, and
thus to be the right procedure of quantization. For more comments in this
precise case of a composite Wess-Zumino field, we also
refer the reader to \cite{BellonMachet2,Machet2,BellonMachet3}.

We quantize the theory by the functional integral formalism. The scalars
and the fermions are not independent degrees of freedom; so, to integrate on
both, we include in the generating functional constraints
\be\ba{l}
\prod_x \delta (H -{v\over \mu^3} \ol\Psi\Psi)(x),\cr
\prod_x \delta (\varphi + i{v\over \mu^3} \ol\Psi\gamma_5{\Bbb T}\Psi)(x),
\ea\ee
that we exponentiate into the effective Lagrangian
\be
{\cal L}_c = \lim_{\beta\rar 0} -{\Lambda^2 \over 2\beta}
\bigg(H^2 + \varphi^2 -2{v\over\mu^3}(H\ol\Psi\Psi -
i\varphi\ol\Psi\gamma_5{\Bbb
T}_L\Psi) + {v^2\over\mu^6}\Big((\ol\Psi\Psi)^2
-(\ol\Psi\gamma_5{\Bbb T}_L\Psi)^2\Big)\bigg).\label{eq:Lcons}
\ee
$\Lambda$ is an arbitrary mass scale. We thus define the theory by
\be
 Z= \int {\cal D} \Psi{ \cal D} \overline\Psi{\cal D}H
{\cal D} \varphi {\cal D} \sigma_\mu \
e^{i\int d^4 x \big({\cal L}(x)+{\cal L}_c(x)\big)} ,\label{eq:Z}
\ee
eventually adding a gauge fixing term. (Remark: the integration over $\tilde
H$ and $\xi$ may be preferred to that over $H$ and $\varphi$, since
${\cal D}\xi$ can be interpreted as the measure over the gauge group
\cite{BabelonSchaposnikViallet}; they only differ by a $\xi$-independent
Jacobian \cite{Machet2}).

${\cal L}_c$ introduces\l
\quad - an infinite bare fermion mass, appearing when $\la H\ra = v$:
\be
m_0 = - {\Lambda^2 v^2\over \beta\mu^3};
\ee
\quad - infinite 4-fermions couplings
\be
\zeta_0 = - \zeta_0^5 = {m_0\over 2\mu^3}.
\ee
At the classical level, the infinite fermion mass in ${\cal L}_c$
is cancelled by the 4-fermions term $\propto(\ol\Psi\Psi)^2$ when
$\la\ol\Psi\Psi\ra = \mu^3$; however, staying in the `Nambu-Jona-Lasinio
approximation' \cite{NambuJonaLasinio}, equivalent to keeping only diagrams
leading in an expansion in powers of $1/N$, the fermion mass and the effective
4-fermions coupling satisfy the two coupled equations
\bea
\zeta(q^2) &=& \frac{\zeta_0}{1-\zeta_0 \; A(q^2, m)},\cr
\noalign{\vskip 2mm}
m &=& m_0 -2\zeta(0)\mu^3,
\eea
graphically depicted in fig.~1 and fig.~2. $A(q^2,m)$ is the one-loop
fermionic bubble. The above cancellation represents only the first two terms
of the series depicted in fig.~2.

\figskip
\epsffile{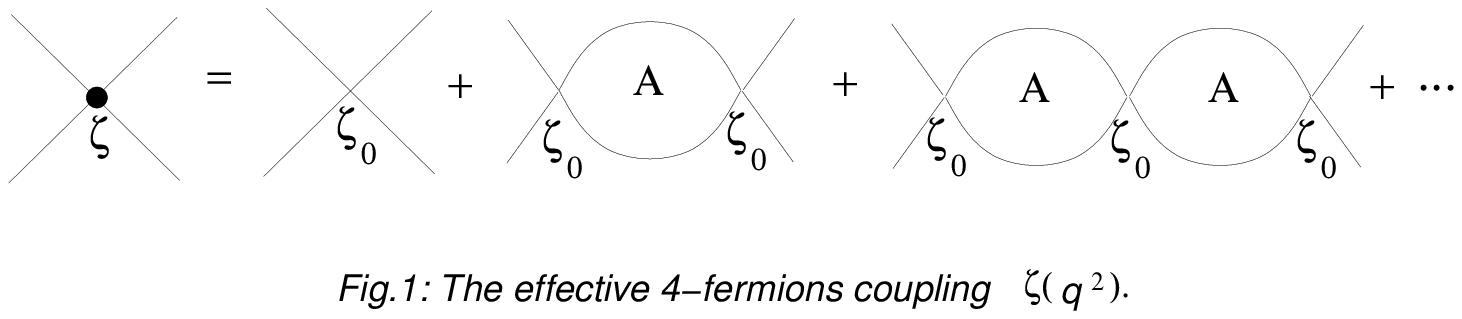}
\figskip
\hskip -.5cm\epsffile{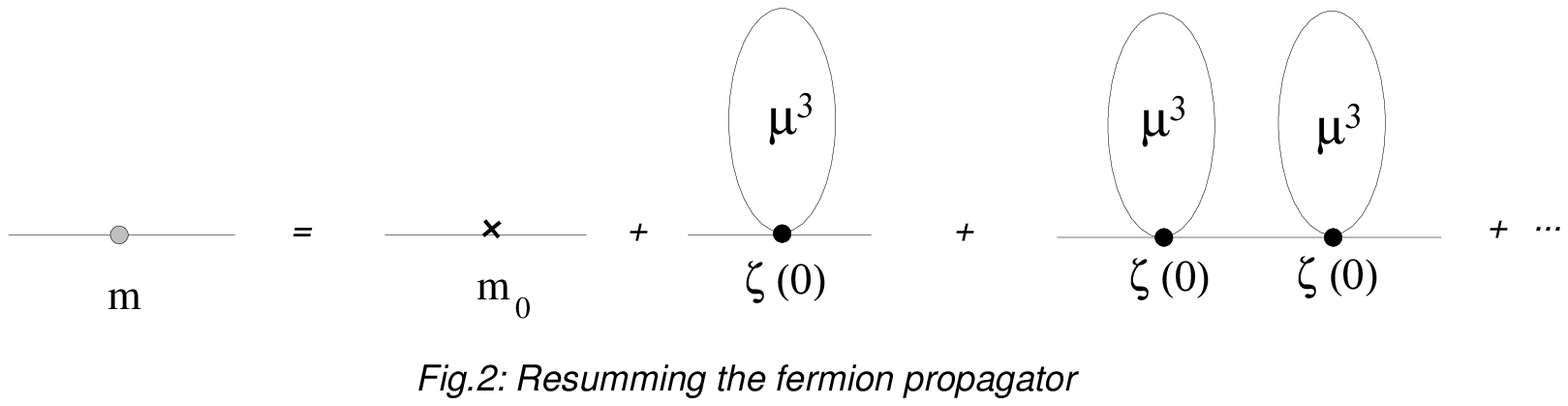}
\figskip
$\mu^3$ being finite, $m=m_0$ is a solution of the equations above
 as soon as $\zeta(0)$ goes to $0$.
This is the case here since $\zeta(0)\propto -A(0,m)^{-1}$, and $A$ involves
a term proportional to $m^2$ (see for example \cite{Broadhurst}). (The
presence of eventual other solutions is beyond our reach because it requires
knowing exactly $A(q^2, m)$).
This also makes the effective 4-fermions coupling $\zeta(q^2)$
(and similarly $\zeta^5(q^2)$) go to $0$ like $\beta^2$.

The fact that the fermions have an infinite mass, in addition to making them
unobservable as asymptotic states (see \cite{Machet2,BellonMachet2})
makes the theory anomaly-free. The Pauli-Villars regularization of the
triangular diagram, which gives the (covariant) anomaly, writes, $M$ being the
mass of the regulator (see fig.~3)
\be
k^\mu \Big(T_{\mu\nu\rho}(m) - T_{\mu\nu\rho}(M)\Big) = m T_{\nu\rho}(m) -M
T_{\nu\rho}(M).
\label{eq:Ward}\ee
We have
\be
\lim _{M\rar\infty} M T_{\nu\rho}(M) = -{\cal A}(g,\sigma_\mu),
\ee
where ${\cal A}(g,\sigma_\mu)$ is the anomaly; so,  when $m\rar\infty$,
the Ward Identity (\ref{eq:Ward}) now shows that the anomaly gets cancelled.
\figskip
\hskip 2cm\epsffile{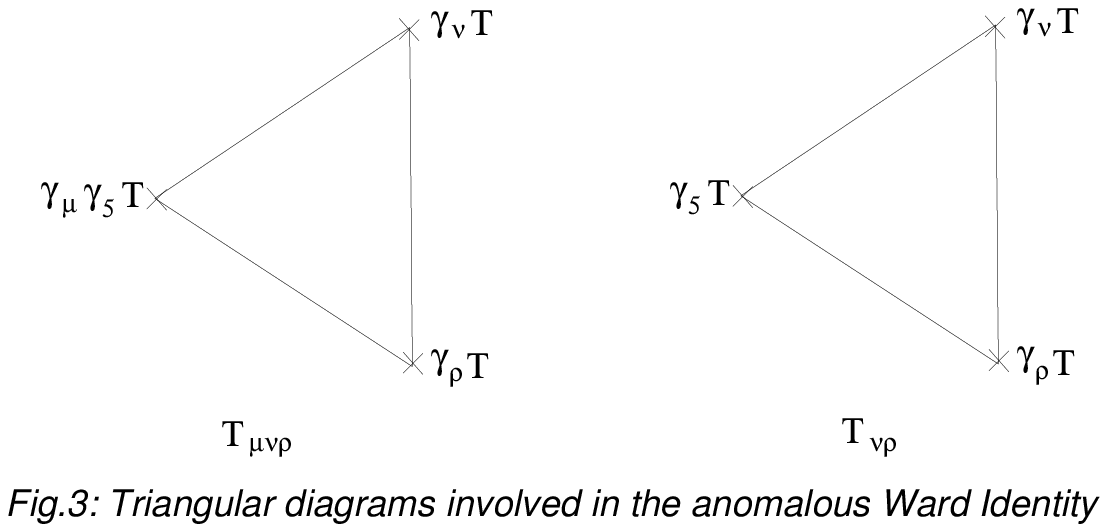}
\figskip
The divergence of the fermionic current also receives a contribution from
${\cal L}_c$ eq. (\ref{eq:Lcons}):
\be
\p_\mu J^\mu_{\psi c} = i\, {\Lambda^2 v\over\beta\mu^3}
(H\ol\Psi\gamma_5{\Bbb T}\Psi -i\,\varphi\ol\Psi\Psi)\; ;
\ee
it can however be consistently taken as vanishing when the constraints hold.

The conservation of the current appearing in the derivative coupling
(\ref{eq:xicpl}) and the vanishing of the effective 4-fermions coupling
constants are arguments in favour of the renormalizability of the theory;
they are developed in \cite{BellonMachet2,Machet2,BellonMachet3}.

\section{Leptonic coupling of the pseudoscalar meson}

The S-matrix element linking a pseudoscalar meson to two leptons is usually
computed by using an effective $current\ \times\ current$ Fermi interaction,
saturating by the vacuum, and using the so-called `Partially Conserved Axial
Current' (PCAC) approximation (see for example \cite{AdlerDashenLee}).
We show here that this PCAC
contribution gets exactly cancelled and that the only contribution to
$\varphi$ into leptons comes from the leptonic equivalent of the coupling
(\ref{eq:xicpl}). Finally we show that a rescaling of the fields allows to
identify it  with the leptonic coupling of a usual (neutral) pion.

We introduce the leptonic multiplet $\Psi_\ell$. There are $N_\ell$ leptonic
flavours. We take here for granted the $V-A$ structure of the weak leptonic
currents. Its origin is studied in \cite{BellonMachet1}.
The leptonic Lagrangian is written
\be
{\cal L}_\ell = i\ol\Psi_\ell \gamma^\mu\Big(\p_\mu
-ig\; (\sigma_\mu -{1\over g}{\p_\mu\xi\over v})\; {\Bbb T}_L\Big)\Psi_\ell\,,
\label{eq:Llept}\ee
where ${\Bbb T}_L$ is now an $N_\ell\times N_\ell$ unit matrix.
We call $L_\mu$ and $H_\mu$ respectively the leptonic and hadronic currents
\bea  H_\mu &=& \ol\Psi \gamma_\mu{\Bbb T}_L \Psi,\cr
      L_\mu &=& \ol\Psi_\ell \gamma_\mu{\Bbb T}_L \Psi_\ell.
\eea
The equation for $\varphi$ is
\be
D^2 \varphi = -\,{1\over v} \p_\mu\; (L^\mu + H^\mu)\, (1-{h\over v})+\cdots,
\ee
giving, neglecting corrections of order $h/v$,
\be
\la 0 \vert \p_\mu (L^\mu + H^\mu)(0) \vert \varphi(k)\ra_{in} =
-v \la 0\vert (\p^2\varphi + gv\,\p_\mu\sigma^\mu +2g\,\sigma_\mu\p^\mu H
-g^2\varphi\,\sigma_\mu ^2) (0)\vert \varphi(k)\ra _{in}.
\ee
The $\p^2\varphi$ term yields the customary `PCAC' equation
\be
\la 0 \vert \p_\mu(L^\mu + H^\mu)(0) \vert \varphi(k)\ra_{in} = vk^2;
\ee
it is however  cancelled by the contribution of $gv\p_\mu\sigma^\mu$, due to
the coupling $gv\,\sigma_\mu\p^\mu\varphi$ occuring in $1/2
D_\mu\varphi D^\mu\varphi$ when, in the low energy regime, one takes the
propagator of $\sigma_\mu$ as $ig_{\mu\nu}/M_\sigma^2$ ($M_\sigma$ is the
mass of the gauge field). This low energy approximation is precisely that
used in the usual `PCAC' computation of the S-matrix element for the
leptonic decay of a pseudoscalar meson, symbolically depicted in fig.~4;
the `bubble' stands for the propagator of the current $L_\mu + H_\mu$,
linked by PCAC to that of the fermionic bound state, and the dot with the
effective Fermi interaction.
\figskip
\hskip 2cm\epsffile{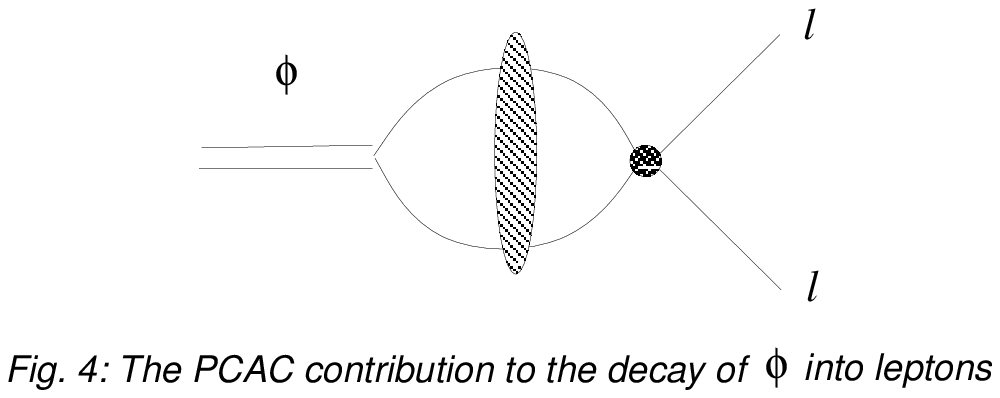}
\figskip
We just found that it is now exactly cancelled by the diagram of fig.~5.
\figskip
\hskip 1cm\epsffile{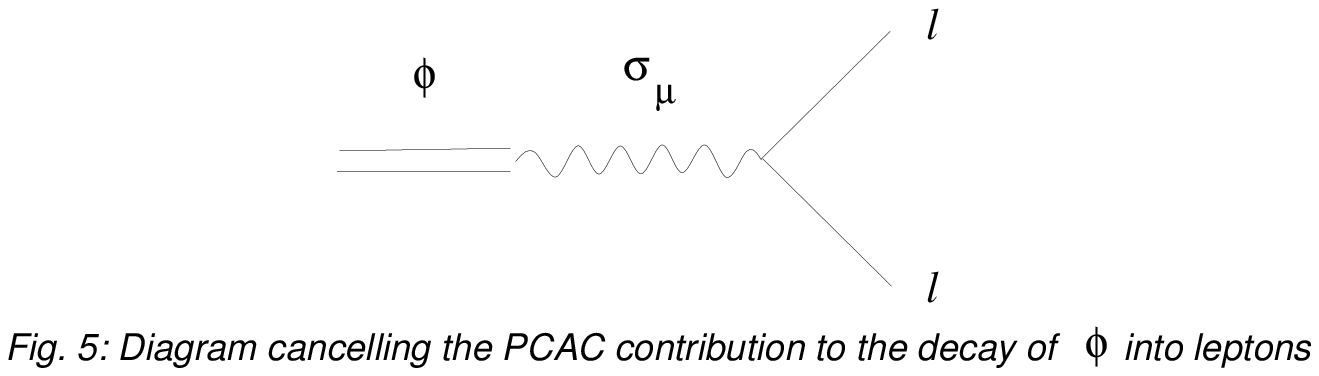}
\figskip
Would the only derivative coupling of $\varphi$ be to the hadronic current,
the same cancellation as above would hold, with the `bubble' of fig.~4 now
standing for the propagation of $H_\mu$ only.
We conclude that, all other contributions cancelling, the disintegrations of
$\varphi$ into leptons are mediated by the direct coupling
$(\xi/v)\,\p_\mu L^\mu$ in (\ref{eq:Llept}).

We rescale the fields by
\bea
\varphi &=& a\pi,\cr
H &=& a H',\cr
\Psi &=& a\Psi',\cr
\Psi_\ell &=& a\Psi'_\ell,\cr
\sigma_\mu &=& a\, a_\mu,\cr
g &=& e/a.
\label{eq:scale}\eea

After a global  rescaling by $1/a^2$, the Lagrangian rewrites (we do not
mention any longer the scalar potential completely `screened' by the
exponentiated constraints)
\be\ba{lcl}
{\displaystyle {1\over a^2}}({\cal L}+{\cal L}_\ell) &=&
-{1\over 4}f_{\mu\nu}f^{\mu\nu}\cr
& & + i\,\ol\Psi' \gamma^\mu(\p_\mu -ie\,a_\mu {\Bbb T}_L)\Psi'
+i\,\ol\Psi'_\ell \gamma^\mu(\p_\mu -ie\,a_\mu {\Bbb T}_L)\Psi'_\ell\cr
& & +{1\over 2}\Big((\p_\mu H' -e \,a_\mu \pi)^2 +
(\p_\mu \pi +e\, a_\mu H')^2\Big)\cr
& & -{a\over v}\;\ol\Psi'_\ell \gamma_\mu {\Bbb T}_L \Psi'_\ell\;\p^\mu \pi,
\ea\ee
where
\be
f_{\mu\nu} = \p_\mu a_\nu - \p_\nu a_\mu.
\ee
We have
\be
\la H'\ra = {v\over a},\ \la\ol\Psi' \Psi'\ra = {\mu^3\over a^2},
\ee
and
\be
e^2 \la H'\ra ^2 = g^2 \la H \ra ^2,
\ee
yielding the same mass $M_\sigma$ for $a_\mu$ and $\sigma_\mu$.
 We call  now $a_\mu$ ``vector boson'', $\pi$ ``pion'',
identify the `primed' leptons with the observed ones and $e$ with the
coupling constant of the theory.
Then, in $({\cal L} +{\cal L}_\ell)/a^2$, the term
\be
-{a\over  v} \;
\ol\Psi'_\ell \gamma_\mu {\Bbb T}_L \Psi'_\ell\;  \p^\mu \pi
\ee
rebuilds the correct S-matrix element for the decay of the pion into two
leptons if we take
\be
a = {f_\pi\over v}.
\ee

\section{The coupling of the pseudoscalar meson to two gauge fields}

The $\varphi$ into two $\sigma_\mu$'s transitions are triggered by the coupling
of ${\cal L}_c$
\be
{i\varphi\over v} m \ol\Psi\gamma_5{\Bbb T}\Psi.
\label{eq:phicpl}\ee
Indeed, the quantum contribution to $m \ol\Psi\gamma_5{\Bbb T}\Psi$
from the triangle precisely yields, as described in (\ref{eq:Ward}) above,
$-i\times\  the\  anomaly$, such (\ref{eq:phicpl}) that contributes at the
one-loop level
\be
{\varphi\over v} {\cal A}(g,\sigma_\mu).
\label{eq:phiano}\ee
Now, after the rescaling (\ref{eq:scale}), (\ref{eq:phiano}) describes
the customary `anomalous' coupling of a neutral pion to two gauge fields:
indeed, we have
\be
{\cal A}(g,\sigma_\mu) = {\cal A}(e,a_\mu);
\ee
consequently, in ${\cal L}_c/a^2$, (\ref{eq:phiano}) becomes
\be
{1\over av}\pi\; {\cal A}(e,a_\mu) = {1\over f_\pi}\pi\; {\cal A}(e,a_\mu).
\label{eq:piano}\ee
Despite the absence of anomaly, it has been rebuilt from the constraints
and the infinite fermion mass that they yield. The fact that the `photon' is
massive is only a formal issue; it is shown in \cite{BellonMachet2} that we
indeed recover the pion decay into two massless photons when they are present.

\section{Conclusion; perspectives}

We emphasized in this work two phenomenological aspects of our model:
we showed that predicting the leptonic couplings of the pion is
compatible with dynamical symmetry breaking and does not necessitate the
introduction of a new scale of interaction nor that of very massive particles;
we also showed that it couples to two gauge fields despite the absence of
anomaly in the hadronic sector with the same strength as usual. The scaling
factor $a=f_\pi/v$ allows the recovery of both in a model where they were
{\em a priori} not expected. This can be traced into the presence of the
coupling (\ref{eq:xicpl}), dictated by gauge invariance
\cite{BabelonSchaposnikViallet}, and into the breaking of the symmetry allowing
the presence of a field $\xi$ with the dimension of a mass and of a coupling
constant $1/v$ with dimension $mass^{-1}$.  We recall that
the leptonic sector has been reconciled in \cite{BellonMachet1} with a
purely vectorial theory and is also itself anomaly-free. The hadronic and
leptonic sectors can consequently be disconnected. The issues of
gauge invariance and unitarity, linked in particular to the
introduction of a derivative coupling between the scalars and the fermionic
current are studied in \cite{Machet2,BellonMachet2}. We show there how the
pion can be gauged into the third polarization of the massive gauge field.
The question of the renormalizability is also developed
 and arguments are given in favour of it; a demonstration at
all orders, going beyond the `Nambu-Jona-Lasinio approximation'
requires a careful study of how the BRS symmetry
\cite{BecchiRouetStora} is implemented in this precise case.
This is currently under investigation \cite{BellonMachet3}.

\bigskip
{\em\underline {Acknowledgements}}: warm thanks are due to M. Bellon for many
suggestions and advice, and his unequalled patience.

\newpage\null
\listoffigures
\bigskip
\begin{em}
Fig.~1: The effective fermion coupling $\zeta(q^2)$;\l
Fig.~2: Resumming the fermion propagator;\l
Fig.~3: The triangular diagrams involved in the anomalous Ward Identity;\l
Fig.~4: The PCAC contribution to the decay of $\varphi$ into leptons.\l
Fig.~5: Diagram cancelling the PCAC contribution to the decay of $\varphi$ into
leptons.\l

\newpage

\end{em}


\begin{thebibliography}{50}
%
\bibitem{GlashowSalamWeinberg}
        S. L. GLASHOW: Nucl. Phys. 22 (1961) 579;\l
        A. SALAM: in ``Elementary Particle Theory:
             Relativistic Groups and Analyticity'' (Nobel symposium No 8),
             edited by N. Svartholm (Almquist and Wiksell, Stockholm, 1968);\l
        S. WEINBERG: Phys. Rev. Lett. 19 (1967) 1224.

\bibitem{AdlerBellJackiwBardeen}
        S. L. ADLER: Phys. Rev. 177 (1969) 2426;\l
        J. S. BELL and R. JACKIW: Nuovo Cimento 60 (1969) 47;\l
        W. A. BARDEEN: Phys. Rev. 184 (1969) 1848.

\bibitem{BouchiatIliopoulosMeyer}
        C. BOUCHIAT, J. ILIOPOULOS and Ph. MEYER: Phys. Lett. 38B (1972) 519;\l
        D. GROSS and R. JACKIW: Phys. Rev. D6 (1972) 477.

\bibitem{Adler}see for example:\l
        S. L. ADLER: ``Perturbation Theory Anomalies'', in ``1970 Brandeis
              University Summer Institute in Theoretical Physics'', vol. 1,
              (Deser, Grisaru and Pendleton eds., M.I.T. Press, 1970).

\bibitem{SusskindWeinberg}
         L. SUSSKIND: Phys. Rev. D20 (1979) 2619;\l
         S. WEINBERG: Phys. Rev. D13 (1975) 974, ibidem D19 (1979) 1277.

\bibitem{FarhiJackiw}see for example:\l
         E. FARHI and R. JACKIW: ''Dynamical Gauge Symmetry Breaking:
            a Collection of Reprints'', (World Scientific, 1982), and
            references therein.

\bibitem{Wess}
         J. WESS: Wigner Medal Acceptance Speach (Salamanca, Spain, 1992).

\bibitem{BellonMachet2}
         M. BELLON and B. MACHET: ``A Not So Standard Model for Quarks'',
                 preprint PAR-LPTHE 1993 to appear.

\bibitem{Machet2}
         B. MACHET: ``Spontaneously Broken $U(1)_L$ Gauge Theories Without
             Asymptotic Fermions and Scalar Boson in 4 Dimensions'',
             preprint PAR-LPTHE (1993) to appear.

\bibitem{BellonMachet3}
         M. BELLON and B. MACHET: in preparation.

\bibitem{BellonMachet1}
         M. BELLON and B. MACHET: ``The Standard Model of Leptons as a Purely
             Vectorial Theory'', preprint PAR-LPTHE 92/18 (revised and
             augmented version, May 1993), hep-ph/9305212.

\bibitem{WessZumino}
        J. WESS and B. ZUMINO: Phys. Lett. 37B (1971) 95.

\bibitem{AbersLee}
         E. S. ABERS and B. W. LEE: Phys. Rep. 9C (1973) 1.

\bibitem{BabelonSchaposnikViallet}
         O. BABELON, F. A. SCHAPOSNIK and C. M.  VIALLET,
                                       Phys. Lett. B177 (1986) 385;\l
         K. HARADA and I. TSUTSUI: Phys. Lett. B183 (1987) 311.

\bibitem{NambuJonaLasinio}
         Y. NAMBU and G. JONA-LASINIO: Phys. Rev. 122 (1961) 345.

\bibitem{Broadhurst} D. J. BROADHURST: Phys. Lett. 101B (1981) 423.

\bibitem{AdlerDashenLee}
         S. L. ADLER and R. F. DASHEN: ``Current Algebra and Application
                      to Particle Physics'', (Benjamin, 1968);\l
         B. W. LEE: ``Chiral Dynamics'', (Gordon Breach, 1972).

\bibitem{BecchiRouetStora}
         C. BECCHI, A. ROUET and R. STORA: Ann. Phys. (N.Y.) 98 (1976) 287.

\end{thebibliography}
\end{document}